# ELECTROENCEPHALOGRAPHIC SLOWING: A PRIMARY SOURCE OF ERROR IN AUTOMATIC SEIZURE DETECTION

*E. von Weltin, T. Ahsan, V. Shah, D. Jamshed, M. Golmohammadi, I. Obeid and J. Picone*

Neural Engineering Data Consortium, Temple University, Philadelphia, Pennsylvania, USA
{eva.vonweltin, tameem.ahsan, vinit.shah, dawer.jamshed, meysam, picone}@temple.edu

*Abstract*— Although a seizure event represents a major deviation from a baseline electroencephalographic signal, there are features of seizure morphology that can be seen in non-epileptic portions of the record. A transient decrease in frequency, referred to as slowing, is a generally abnormal but not necessarily epileptic EEG variant. Seizure termination is often associated with a period of slowing between the period of peak amplitude and frequency of the seizure and the return to baseline. In annotation of seizure events in the TUH EEG Seizure Corpus, independent slowing events were identified as a major source of false alarm error. Preliminary results demonstrated the difficulty in automatic differentiation between seizure events and independent slowing events. The TUH EEG Slowing database, a subset of the TUH EEG Corpus, was created, and is introduced here, to aid in the development of a seizure detection tool that can differentiate between slowing at the end of a seizure and an independent non-seizure slowing event. The corpus contains 100 10-second samples each of background, slowing, and seizure events. Preliminary experiments show that 77% sensitivity can be achieved in seizure detection using models trained on all three sample types compared to 43% sensitivity with only seizure and background samples.

## I. INTRODUCTION

An electroencephalographic (EEG) record, especially an abnormal one, can exhibit features associated with seizures without containing a seizure [1]. Brief evolution of epileptiform discharge, focal and generalized slowing, and movement artifacts such as chewing are all common features of seizures that can be observed in non-ictal (non-seizure) portions of the record [2]. EEG features that resemble seizures, as shown in Figure 1, or resemble pre-ictal and post-ictal portions occurring directly before or immediately following a seizure pose a challenge to seizure detection studies. Unsurprisingly, these features are among the top identified causes for false alarm errors in automatic seizure detection.

A human reviewing an EEG record can differentiate between instances of features that coincide with seizure events versus instances not associated with seizures based on events preceding and following the event. Automatic systems must be able to differentiate between these events based on the morphologies of features associated with and separate from seizure events. Profound muscle artifacts, for example, can completely obscure the record. A human looking for a seizure would examine the signal immediately preceding the muscle artifact for small indications of epileptic evolution such as a gradual increase in frequency and amplitude of spike and slow wave complexes. Typical indicators of a seizure, such as the epileptic evolution, might only be visible for a few seconds before disappearing under the muscle artifact. The seizure persists during the artifact, leading to the artifact itself being annotated as part of the seizure event. An automatic seizure detection system trained using these data then erroneously detects muscle artifact not associated with an epileptic event as a seizure.

Profound muscle artifacts [3] are a common example of sources of false alarm errors in seizure detection. These false alarms, however, represent a less complex problem than false alarms on slowing events. Muscle artifacts, unlike electrographic slowing, are not produced in the brain. Rather than training a system to differentiate between muscle artifact obscuring a seizure and muscle artifact occurring independently, muscle artifacts are typically just removed through artifact reduction [4]. Slowing however, can be part of a seizure, and should be detected as such. To achieve this, the system must be able to differentiate between slowing at the end of a seizure and slowing that occurs independent of a seizure event. As shown in Figure 1, without the context of whether there was a seizure event directly preceding it, slowing at the end of a seizure, referred to as post-ictal slowing, and independent slowing events can look very similar.

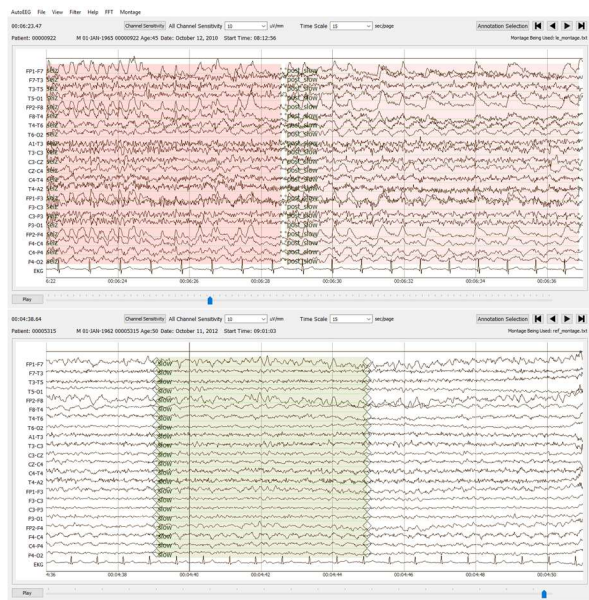

Figure 1. Example of a seizure (left annotation in upper image) followed by an example of post-ictal slowing (right annotation in upper image) and an example of a non-seizure slowing event that resembles post-ictal slowing (lower image).

While slowing at the end of a seizure is generally understood to be a sign of the brain's gradual return to baseline following an epileptic event, the pathogenesis of independent slowing events can be unclear. EEG slowing can be seen in patients with cerebral structural abnormalities, during interictal periods in epileptic patients, and even during normal sleep [5]. Interictal regional delta slowing (IRDS) also represents a challenge to seizure detection. Often present in temporal proximity to seizures, IRDS can be rhythmic and sharply contoured, closely resembling post-ictal slowing.

Focal and generalized EEG slowing was found to be associated with a significant portion of false alarm events in EEG records annotated as part of the Temple University Hospital Seizure Detection Corpus (TUSZ) [6]. Delta waves at the end of and following seizures are a common feature of seizure termination [7]. There are numerous variants in seizure morphology, evolution, and termination. The seizure time course, from the initial evolution of the seizure to its termination, generally consists of well-defined pre-ictal, ictal, and post-ictal stages. Seizures begin with evolution in amplitude and/or frequency. Following the maximal ictal period, slowing can be observed. In annotating seizures, an effort is made to capture the entire time course of the event, including post-ictal slowing, as the progression in frequency and amplitude is what really differentiates the event from baseline. Slowing at the end of and directly following a seizure therefore ends up as part of the seizure annotation. This leads to the erroneous detection of independent slowing events as seizures.

Slowing at the end of seizures plays a very important role in describing the termination of a seizure and should not be simply excluded from seizure annotations. Instead, the Temple University Hospital Slowing Corpus (TUSL), introduced here, was created to aid in the development of an automatic system that can differentiate between post-ictal and transient slowing.

## II. THE TUH EEG SLOWING CORPUS

TUSL is an annotated subset of the TUH EEG Corpus (TUH-EEG) [8], the world's largest publicly available database of clinical EEG data. It currently includes over 30,000 sessions, over 16,000 patients and 29 years of signal data. The files from which examples of slowing, seizure, and complex background were selected are part of the TUSZ training set.

Slowing events were annotated by undergraduate students at Temple University. These data annotators have been evaluated on the quality and consistency of their annotation through inter-rater agreement studies and shown to perform with sufficiently high accuracy [9]. Students were trained previously to identify and annotate seizure events for use in the training and evaluation of automatic seizure detection. Though they had been trained to differentiate between seizure events and slowing events, they had not been trained to identify and annotate independent slowing events. Therefore, standards for slowing annotations were developed.

In Figure 2 we show several examples of slowing that were included in the corpus. These demonstrate focal and generalized slowing, which are often associated with diffuse and localized brain dysfunction, respectively. Slow wave activity can be seen in patients with and without epilepsy as brain lesions from any pathological source can result in deafferentation [10] of cortical neurons. Deafferentation, the loss of afferent, or input, connections on cortical neurons may be responsible for slowing events both local and diffuse. Intermittent slowing events vary in duration but generally last less than 10 seconds.

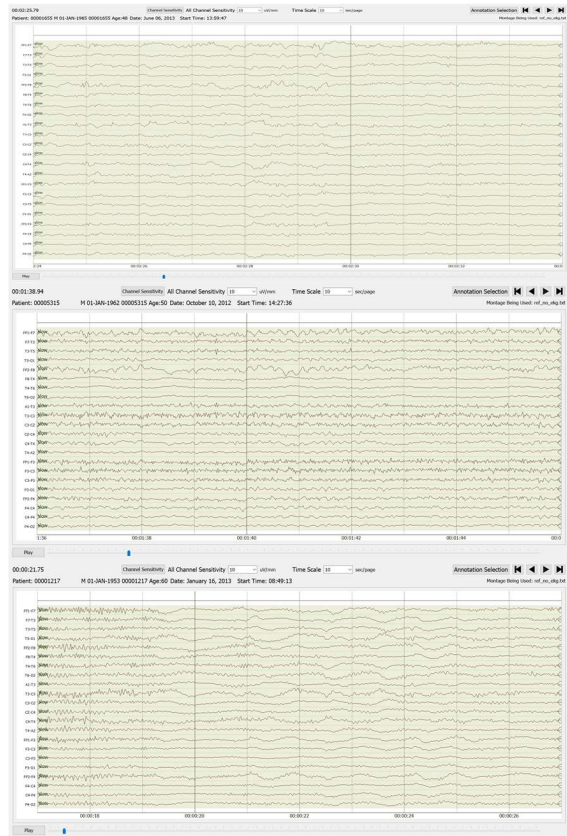

Figure 2. Electroencephalographic slowing can present in many forms. In these examples, the slowing event is centered in the 10-second sample. Somewhat sharply contoured generalized slowing (upper image); non-generalized, low amplitude slowing (middle image); and slowing with very smooth, slow waves (lower image) can all be found in the corpus.

The challenge in annotating slowing events arises from the great variety of what is widely considered to be electrographic slowing [5]. Background slowing and continuous slowing were not annotated as they are not associated, in our experience, with false alarms in machine learning experiments. Background slowing refers to a slowing of the normal posterior dominant rhythm seen in awake EEG recordings. Continuous slowing has not been included as a feature of slowing in this dataset as it does not constitute a clear change from background, and this change from background is what is associated with false alarms. Intermittent, generalized slowing was the focus of these annotations as it most closely resembles the slowing at the end of seizures.

We identified 100 10-second examples of slowing. Within the 10-second sample, a slowing event may last 2-10 seconds. Efforts were made to ensure that these 100 samples do not contain seizures or artifacts such as muscle, lead, or chewing artifacts. Slowing samples were selected from files not containing seizures.

We also identified 100 10-second samples of seizure events for comparison. Seizure samples were generated from seizure annotations in TUSZ. Seizures longer than 10 seconds in duration were randomly selected from these files. Ten second samples were then taken from the midpoint of the seizure event in the annotation. In this way, a variety of both seizure morphologies and stages of evolution, from seizure initiation to termination, were captured.

We also gathered 100 10-second samples of complex background, again using the TUSZ training set. Slowing events, though not considered seizures, were considered separate from background in this corpus. Background samples were taken from the same sessions from which slowing and seizure samples were selected, though no file contains both seizure and slowing samples. Patients with seizures often never actually experience what could be considered clinically normal background [2]. For this reason, background samples were selected to contain complex features such as muscle, chewing, and lead artifacts. Annotators were careful to not include any examples of slowing in the background samples.

In total, 112 files were annotated for use in this dataset: 61 files contain seizure samples, with 18 of these also containing background samples; 45 files contain slowing samples, with 26 of these also containing background files. The remaining 6 files contain only background samples. These 6 files were taken from the same sessions that the seizure and slowing files are from. Figure 3 contains exemplary samples of each annotation type.

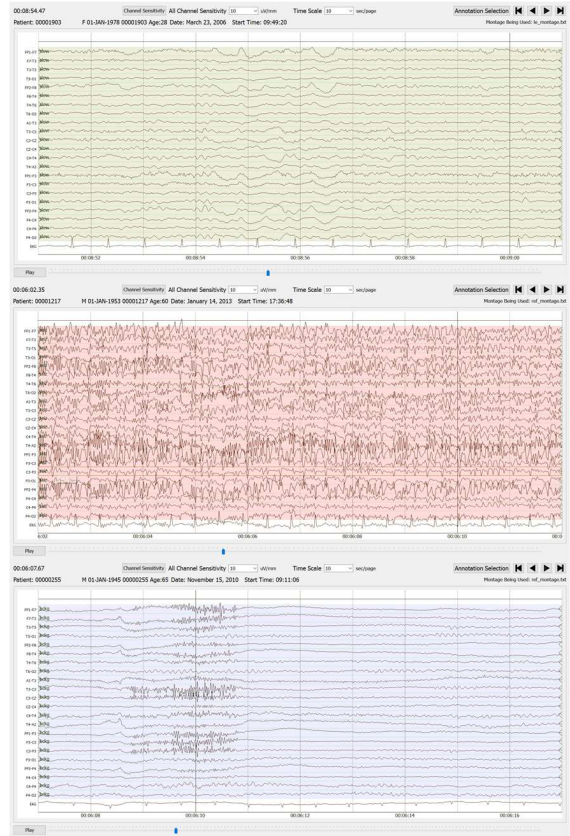

Figure 3. Examples of 10-second samples: intermittent, generalized slowing sample (upper image); seizure sample at its point of maximum amplitude and frequency (middle image); and complex background sample containing muscle artifact (lower image).

### III. PRELIMINARY EXPERIMENTS

Our goal in developing this corpus was to augment the training process for our machine learning systems [11]. Our first effort to develop solutions to the problem of a high false alarm rate on independent slowing events was to identify 30 files that do not contain seizures but do contain intermittent slowing. We then augmented the training database with this data and retrained the system. Unfortunately, the system failed to train. Suspicious that this was a problem of regularization, we conducted an experiment in which Gaussian Noise layers [12], regularization layers meant to keep models from overtraining, were removed. Following the change, the system was now able to train, though performance decreased compared to training using only files containing seizures. This counterintuitive result was the motivation for developing this corpus.

Three experiments were conducted using this database to initialize training of a seizure detection system: (1) using only the seizure samples; (2) augmenting the seizure samples with the slowing samples; and (3) combining the seizure, slowing, and background samples. The model is first trained on the 10-second sample data, overfit on those data, and then re-trained using the original training set. In all experiments, the system failed to train properly (sensitivity was 0%). These experiments were designed to test the hypothesis that false alarms on slowing events are caused by a failure in training to differentiate between slowing and seizure events.

This database was also used to train models independent of the TUSZ training set. Consistent with our previous findings, regularization was found to be a barrier to successfully training a system. A Convolutional Neural Network (CNN) [11] model trained and evaluated in a closed-loop fashion had an initial error rate of 25%. Error rates were reduced to 20% when preserving dropout layers only in the first and second layers of CNNs. After removing all drop-out layers the error rate reduced to 0%. Drop-out layers are a method of regularization designed to prevent overfitting by randomly selecting units to remove from the neural network [11].

Table 1 presents a summary of the open loop experiments conducted with this corpus. We used a cross-validation approach since there was a limited amount of data. A Multi-Layer Perceptron (MLP) model was trained using 3 different subsets of the 100 10-second samples in this corpus. The model was trained on seizure and complex background files (2-way -slow in Table 1); seizure and slowing files (2-way + slow in Table 1); and on seizure, slowing, and complex background files (3-way in Table 1). All models were evaluated for performance in seizure detection. Before splitting the files into training and evaluation sets the files were either randomly shuffled or were not (denoted 'Shuffling' and 'No Shuffling' in Table 1). Each model was then trained on 80% of the files and evaluated for seizure detection on the remaining 20%. This process was repeated until all files had been in the evaluation set at least once.

Results are a combination of the evaluation of each of these repetitions. In models both with and without shuffling, sensitivity and specificity in seizure detection are improved when the model is trained with the slowing files. Sensitivity is improved by shuffling before sorting into training and evaluation, though specificity is slightly decreased. Only with shuffling do we see better results on the '3-way' model than the '2-way + slow' model. The overall classification performance is slightly superior to more traditional approaches based on random forests.

| Feature | Shuffling | | | No Shuffling | | |
|---|---|---|---|---|---|---|
| | 2-Way | | 3-way | 2-Way | | 3-way |
| | - slow | + slow | | - slow | + slow | |
| Sensitivity | 43.7% | 73.2% | 77.2% | 31.5% | 70.1% | 64.6% |
| Specificity | 87.6% | 93.6% | 92.2% | 84.7% | 94.8% | 95.6% |

Table 1. A summary of preliminary open loop experiments.

IV. SUMMARY

Electrographic seizures, though abnormal, do share morphology with non-seizure EEG features. For successful automatic seizure detection to occur, systems must be able to differentiate between epileptic and non-epileptic variants. Post-ictal slowing seen at seizure termination is morphologically like that of intermittent EEG slowing. Independent slowing events have been found to be responsible for false alarm errors in seizure detection. We have introduced the TUH EEG Slowing Corpus (TUSL) to support investigations into these problems.

This corpus consists of 100 10-second samples of slowing, with the slowing event centered in each 10-second sample. It also contains 100 control conditions for seizures and complex background events. These data are a subset of both the TUH EEG Corpus (TUH-EEG) and the TUH EEG Seizure Corpus (TUSZ) training set.

Preliminary results of experiments with machine learning has yet to show a significant improvement in performance when pre-training using this data. We demonstrated that there are several operational issues yet to be understood, including the role of regularization and randomization in training. Open loop experiments found that training using the slowing annotations improves seizure detection compared to a model trained only with seizure and background files. Training using all three annotation types improved performance compared to a model trained using only slowing and seizure files when using shuffling.

We expect to continue developing resources to improve seizure detection technology. Open source data and resources can be found on our web site (see *https://www.isip.piconepress.com/projects/tuh_eeg/*).

ACKNOWLEDGEMENTS

Research reported in this publication was most recently supported by the National Human Genome Research Institute of the National Institutes of Health under award number U01HG008468. The content is solely the responsibility of the authors and does not necessarily represent the official views of the National Institutes of Health. This material is also based in part upon work supported by the National Science Foundation under Grant No. IIP-1622765. Any opinions, findings, and conclusions or recommendations expressed in this material are those of the author(s) and do not necessarily reflect the views of the National Science Foundation. The TUH EEG Corpus work was funded by (1) the Defense Advanced Research Projects Agency (DARPA) MTO under the auspices of Dr. Doug Weber through the Contract No. D13AP00065, (2) Temple University's College of Engineering and (3) Temple University's Office of the Senior Vice-Provost for Research.